\documentclass[prl,reprint,amsmath,amssymb,nopacs,floatfix]{revtex4-1}

\usepackage[breaklinks=true,colorlinks,citecolor=blue,linkcolor=blue,urlcolor=blue]{hyperref}
\usepackage{epsfig,mathrsfs,color,latexsym,subfigure,marginnote,physics}

\renewcommand{\BibitemShut}[1]{}

\newcommand{\Fig}[1]{Fig.~\ref{#1}}

\renewcommand{\section}[1]{{\par\it #1.---}\ignorespaces}

\begin{document}
\title{Spontaneous ferromagnetism and finite surface energy gap in the topological insulator Bi$_2$Se$_3$ from surface Bi$_\mathrm{Se}$ antisite defects}

\author{Suhas Nahas}
\email[]{suhas.nahas@physics.uu.se}
\author{Biplab Sanyal}
\email[]{biplab.sanyal@physics.uu.se}
\author{Annica M. Black-Schaffer}
\email[]{annica.black-schaffer@physics.uu.se}
\affiliation{Department of Physics and Astronomy, Uppsala University, Box 516, SE-751 20 Uppsala, Sweden}
\date{\today}

\begin{abstract}
We perform ab-initio calculations on Bi$_\mathrm{{Se}}$ antisite defects in the surface of Bi$_2$Se$_3$, finding strong low-energy defect resonances with a spontaneous ferromagnetism, fixed to an out-of-plane orientation due to an exceptional large magnetic anisotropy energy.
For antisite defects in the surface layer, we find semi-itinerant ferromagnetism and strong hybridization with the Dirac surface state, generating a finite energy gap. For deeper lying defects, such hybridization is largely absent, the magnetic moments becomes more localized, and no energy gap is present.
\end{abstract}

\maketitle


     Topological insulators (TI) \cite{Hasan_Kane10,Qi_Zhang11} has been one of the most intensively studied areas in physics in the past decade, owing to their remarkable electronic properties: the bulk is insulating but the surfaces are metallic due to a gapless Dirac surface state (DSS) protected by time reversal symmetry (TRS).
     
     Besides an interest in discovering new TIs, considerable effort has also been dedicated  to opening up an energy gap in the DSS spectrum by breaking TRS, in order to enhance the electric control and also achieve the quantum anomalous Hall effect (QAHE)\cite{Haldane15,Onoda_Nagaosa03,Liu_Zhang08}. A natural route to break TRS is to introduce an effective magnetic field perpendicular to the surface of the TI \cite{Tokura_Tsukazaki19}. Most of the studies along this route have involved doping the TI with magnetic impurities \cite{Hor_Cava10,Yu_Fang10,Chang_Xu13,Chang_Moodera15,Peixoto_Reinert16,Zhang_Wu18,Tokura_Tsukazaki19,Liu_Wang19,Bestwick_Goldhaber-Gordon15,Kou_Kang14}, whose magnetic moments might couple ferromagnetically through RKKY \cite{Abanin_Pesin11,Zhu_Chang11}, Van-Vleck \cite{Li_Zhu15}, or other exchange \cite{Vergniory_Ernst14,Peixoto_Reinert16,Zhang_Wu18,Ye_Kimura19,Tcakaev_Hinkov20} mechanisms to produce the necessary out-of-plane magnetic field. A more recent development has been the realization of intrinsic magnetism in MnBi$_2$Te$_4$ \cite{Li_Xu19, Zhang_Wang19, Zeugner_Isaeva19, Gong_Feng19}, where the Mn atoms order antiferromagnetically with an out-of-plane magnetic anisotropy. 
     
     However, for both magnetic impurities in TI and MnBi$_2$Te$_4$, there seem to exist significant complications when it comes to opening a gap in the DSS, with experiments so far reporting both the presence of an energy gap \cite{Chang_Xu13,Sanchez-Barriga_Rader16,Otrokov_Chulkov19,Chen_Shen10,Wray_Hazan11}  and finite density of states \cite{Valla_Chu12,Scholz_Rader12,Li_Ding19,Hao_Liu19,Chen_Chen19,Schlenk_Wiesendanger13,Beidenkopf_Yazdani11,Sessi_Bode14,Yang_Jia13,Okada_Madhavan11} at the Dirac point. Also, in the case of thin films, the hybridization between the two DSSs could be the reason for a finite energy gap \cite{Linder_Sudbo09,Lu_Shen10,Liu_Zhang10}, and not TRS breaking.
    For magnetic impurities, a two-fluid description has been proposed \cite{Sessi_Balatsky16} to account for the contradicting results. Here the DSS spectrum is indeed gapped due to TRS breaking, but at the same time the non-magnetic part of the scattering potential produces localized impurity-induced resonances \cite{Biswas_Balatsky10, Black-Schaffer_Balatsky12a,Black-Schaffer_Balatsky12b, Alpichshev_Kapitulnik12, Wehling_Balatsky14} filling up the gap \cite{Black-Schaffer_Balatsky15}. 
  
  In this work, we show that a surface energy gap is generated in the most common TI, Bi$_2$Se$_3$, from intrinsic  Bi$_{\mathrm{Se}}$ antisite defects, entirely without the need of foreign magnetic atoms. By performing extensive ab-initio calculations of antisite defects, we find defect-induced low-energy resonances, which spontaneously acquire a magnetic moment and thus gap the DSS. Antisite defects and their associated resonance states have already been observed experimentally using scanning tunneling microscopy (STM)\cite{Urazhdin_Mahanti02, Kim_Kimura11, Mann_Shih13, Dai_Wu16}, in both surface and subsurface layers, when growing Bi$_2$Se$_3$ in a Bi-rich environment \cite{Zhang_Yao12, West_Zhang12, Scanlon_Catlow12, Xue_Zhong13}. An additional benefit of Bi$_\mathrm{Se}$ defects is that they behave as compensating p-type dopants, neutralizing the naturally occurring n-type Se vacancies by moving the Dirac point closer to the Fermi level \cite{Hor_Cava09, Xue_Zhong13}.

In detail, we show how Bi$_{\mathrm {Se}}$ antisite defects in the TI surface produce low-energy states, with a spontaneous magnetization which even increases for lower concentrations. We find a magnetic anisotropy energy favoring an out-of-plane magnetic orientation of individual antisite defects that is two orders of magnitude larger than for common magnetic dopants. Together with an appreciable ferromagnetic exchange coupling this guarantees an out-of-plane ferromagnetic alignment between different defects. For antisite defects in the surface layer, we find semi-itinerant ferromagnetism and defect states coupling strongly to the DSS, resulting in a sizable energy gap in the DSS. On the other hand, antisite defects in the first subsurface layer display more localized magnetism with no discernible hybridization with the DSSs and consequently no DSS energy gap. This also reveals that a significant hybridization is necessary between the DSS and the  defect states for the magnetic moment to be able to produce an energy gap. Taken together, our results open up an entirely original and general pathway for designing magnetic and gapped TIs, by merely tuning the synthesizing conditions and thus completely avoiding the need for external magnetic impurity atoms.

\section{Method}
\begin{figure}[htb]
\begin{center}
\includegraphics[width=0.80 \linewidth]{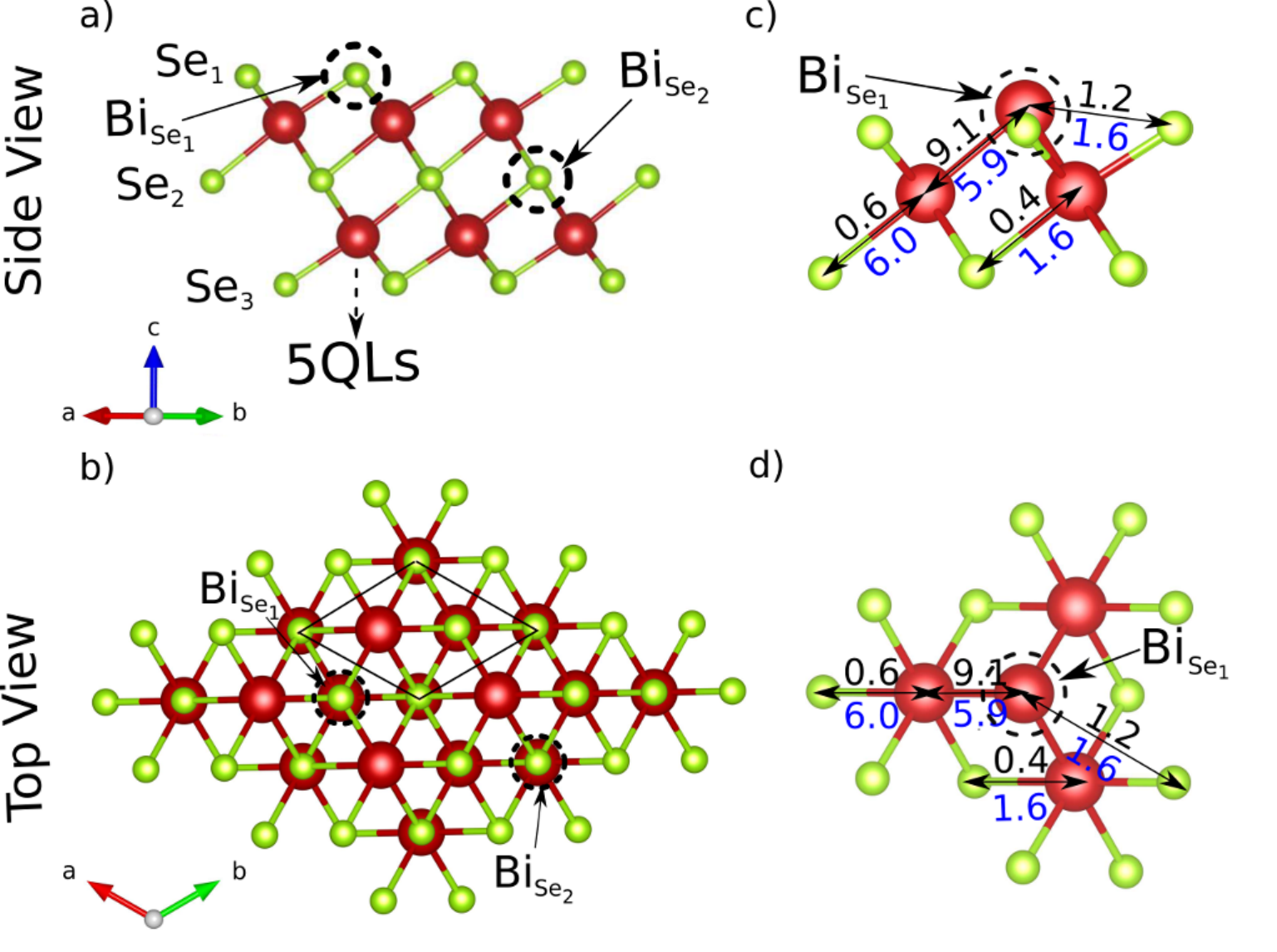}
\caption{Side (a) and top (b) views of the pristine Bi$_2$Se$_3$-slab, with Bi (Se) atoms in red (green) and the conventional lattice vectors of Bi$_2$Se$_3$ used as reference axes. Location of Bi$_\mathrm{Se_1}$ and Bi$_\mathrm{Se_2}$ antisite defects are indicated by dotted circles and surface unit-cell of pristine Bi$_2$Se$_3$ by black parallelogram. Side (c) and top (d) views of neighborhood of Bi$_\mathrm{Se_1}$ antisite defect with total relative (in \%) bond length change in the presence (absence) of spin-orbit coupling in black (blue). 
}
\label{f1}
\end{center}
\end{figure}

We perform electronic structure calculations, based on density functional theory, as implemented in the Vienna Ab-initio Simulation Package (VASP \cite{Kresse_Furthmuller96}), on Bi$_2$Se$_3$-slabs containing six  quintuple layers (Se$_1$-Bi$_1$-Se$_2$-Bi$_2$-Se$_3$) in order to capture the TI surface, while still maintaining bulk conditions within the slab. On the surface we create a supercell by repeating the conventional surface unit cell, $n \times n$ ($n = 2,3,4$), adding one defect per supercell, resulting in defect concentrations x $\sim$ 25, 11, and 6\%. Below we mainly report results for antisite defects Bi$_\mathrm{Se_{1,2}}$, i.e.~Bi replacing either the surface Se$_1$ or subsurface Se$_2$ atom, see \Fig{f1}(a,b), but we also study Bi$_\mathrm{Se}$ defects in deeper layers, including the bulk.
We carry out the structural and electronic optimizations using a plane-wave basis set with kinetic energy cut-off 270~eV \footnote{We have checked that increasing the kinetic energy cut-off to 600 eV does not change the results for the  11\% Bi$_\mathrm{Se_{1,2}}$ defect}, together with Projector Augmented Wave (PAW) pseudopotentials. We use the GGA for the exchange-correlation functional \cite{Shirali_Vekhter19} and DFT-D3 \cite{Grimme_Kreig10} to properly account for the van der Waals corrections. Furthermore, we use a $\Gamma$-centered $k \times k \times 1$ grid to sample the Brillouin zone, where for even (odd) $n$ we use  $k \times n = 8 (9)$ and $k \times n = 4 (3)$ for the electronic and structural optimizations, respectively. We also use a 30 \AA\ vacuum to isolate each periodic instance of the slab. In terms of convergence criteria, we use force and energy convergence thresholds of 10$^{-6}$ eV (corresponding to $3 \times 10^{-2}$ meV/$\mathrm{\AA}$) and $10^{-7}$ eV, respectively. We perform all calculations in a \textcolor{black}{scalar} relativistic manner, always including the effects of spin-orbit coupling, and also allow for a finite magnetization in all directions. 

\section{Structural distortions}
We start by performing structural optimizations of the atomic positions for each defected TI surface. This both establish the equilibrium positions of the Bi$_{\mathrm{Se}}$ defects and give a structural view on the impact of antisite defects. To quantify the latter, we track the atomic distortions caused by the Bi$_\mathrm{Se}$ defects by comparing with an equivalently relaxed pristine TI slab. In \Fig{f1}(c,d) we display in black text the relative change of bond lengths (in \%) in the neighborhood of the  Bi$_\mathrm{Se_1}$ defect, while blue text reports the equivalent change when ignoring spin-orbit coupling. 
As seen, the Bi$_\mathrm{Se_1}$ defect creates large local perturbations of the lattice structure, with bond lengths changing as much as 9\% for nearest neighbor bonds. This is by all accounts a large structural change, which we at least partly can attribute to the 40\% larger atomic size of the Bi atom compared to Se. In comparison, the next-nearest  neighbor bonds show almost negligible distortion. If we were to ignore the spin-orbit coupling in the structural optimization we find that both the nearest and next-nearest neighbor bond show a similar change. This illustrates that spin-orbit coupling is essential to capture not just the DSS but also the correct atomic structure of antisite defects in TIs.
We find similar structural patterns for the other antisite defects, including defects in the bulk of the TI, see Supplementary Material (SM) \cite{SM}.

\section{Magnetism}
We next turn to the electronic properties of antisite defects. Surprisingly we find that Bi$_2$Se$_3$ with antisite surface defects hosts a pronounced magnetization, despite the intrinsically non-magnetic nature of antisite defects. For both Bi$_\mathrm{Se_1}$ and Bi$_\mathrm{Se_2}$, we observe a highly anisotropic, out-of-plane (c-direction), magnetization. Bi$_\mathrm{Se_3}$ (Bi on the third Se layer) also gives rise to a net magnetization, but the defect also easily migrates to the van der Waals gap during structural optimization.  
For antisite defects further into the bulk we find no magnetization. Interestingly, if we start with atomic structures optimized {\it without} spin-orbit coupling, we find no net magnetization for antisite defects in any layer, \textcolor{black}{including Bi$_\mathrm{Se_{1,2}}$}. Thus, the large structural distortions caused by spin-orbit coupling is crucial for correctly determining the electronic ground state of antisite defects.
\begin{figure}[]
\begin{center}
\includegraphics[width=0.99 \linewidth]{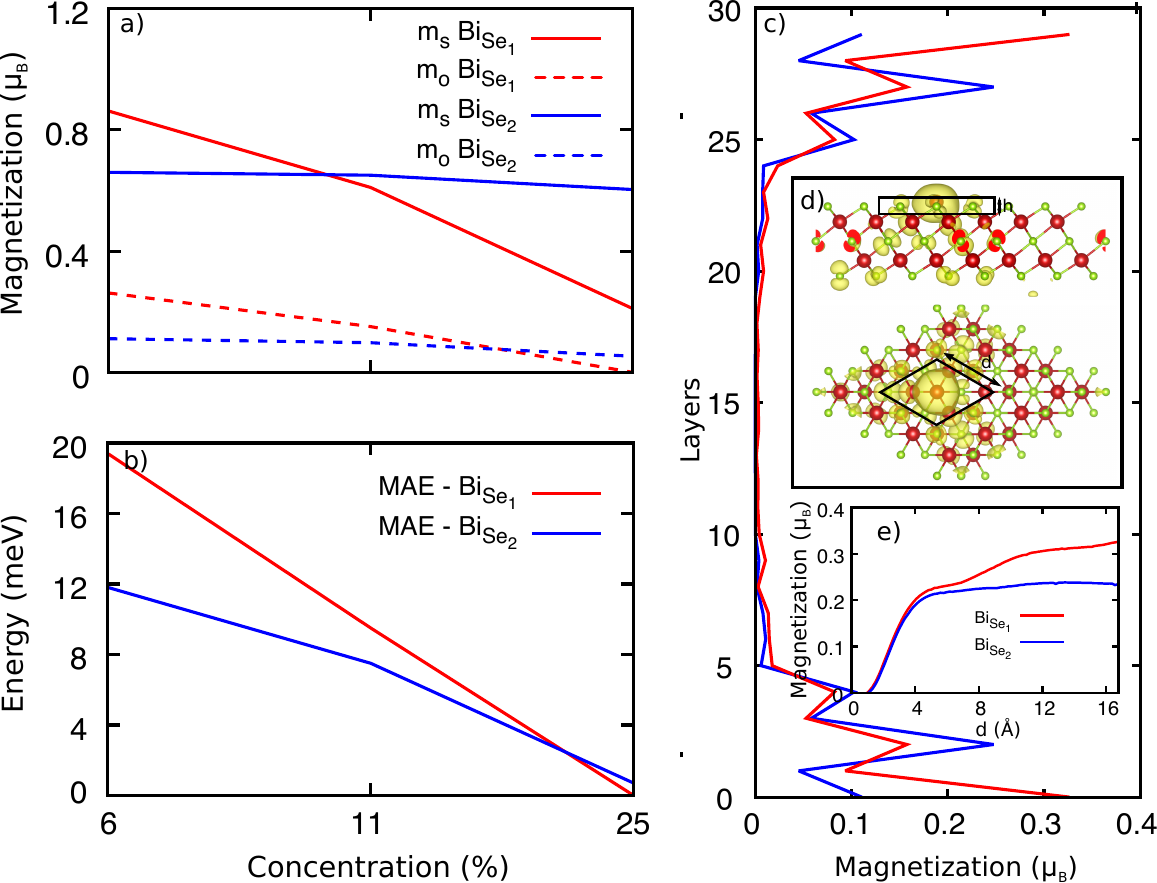}
\caption{(a) Magnetization, spin $m_s$ (solid) and orbital $m_o$ (dashed), and (b) MAE for  Bi$_\mathrm{Se_1}$ (red) and Bi$_\mathrm{Se_2}$ (blue) antisite defects as a function of defect concentration. (c) Layer-resolved spin magnetization for each atomic layer for 6 \% Bi$_\mathrm{Se_1}$ (red) and Bi$_\mathrm{Se_2}$ (blue) antisite defects. (d) Real-space magnetization density in first surface layer for Bi$_\mathrm{Se_1}$ defect with isovalue 0.1 times the maximum. (e) Integrated spin magnetization for Bi$_\mathrm{Se_1}$ (red) and Bi$_\mathrm{Se_2}$ (blue) defects over the volume of the black rhombus displayed in (d) (centered at the antisite defect with side $d$ and thickness $h$), given as a function of $d$ with $h$ equal to the Bi-Se c-axis projected bond-length.
}
\label{f2}
\end{center}
\end{figure}

In Fig.~\ref{f2}(a) we show how the net magnetization varies as a function of defect concentration for the Bi$_{\mathrm{Se_1}}$ and Bi$_{\mathrm{Se_2}}$ defects. We find that both spin and orbital moments increase with decreasing concentration: the surface Bi$_{\mathrm{Se_1}}$ defect shows an almost three-fold increase in the spin magnetic moment when decreasing the defect concentration from 25\% to 6\% (see SM \cite{SM}), while the subsurface Bi$_{\mathrm{Se_2}}$ defect shows a minor increase. The increasing, and persistent, magnetization with decreasing defect concentrations assures that the magnetization is stable in the dilute defect limit. 
In Fig.~\ref{f2}(a) we also see that the orbital moments are large, \textcolor{black}{strongly} suggestive of a highly anisotropic magnetization \cite{Bruno89,Skomski_Enders11,Rau_Brune14,Ou_Wu15}. In order to confirm this, we calculate magnetic anisotropy energy (MAE), i.e.~the total energy difference between out-of-plane and in-plane magnetizations. Figure \ref{f2}(b) shows how also the MAE increases significantly with decreasing defect concentration. Notably, the MAE is almost 20 (12) meV for the Bi$_\mathrm{Se_{1(2)}}$ antisite defect at the lowest concentration. Such MAE values are impressive, about two orders of magnitude larger than what has been achieved in TIs with the magnetic transition metal impurities Cr, V or Mn, where the MAE is only of the order of 0.1 meV \cite{Islam_Sessi18}. 
We also directly calculate the exchange coupling as the energy difference between ferromagnetic and anti-ferromagnetic c-axis alignments of \textcolor{black}{two Bi$_{{\mathrm{Se}}_1}$ defects at a distance of $\sim$ 14.5 \AA} and $\sim 6\%$ concentration, see SM \cite{SM}. We find the ferromagnetic configuration to be lower in energy by \textcolor{black}{3.6 meV. This is a large value considering the distance between the defects and contributes by a large amount ($\sim$ 500 meV${\mathrm{\AA}}^2$)\cite{Pajda01}  to the spin-stiffness as predicted for spontaneous magnetization \cite{Edwards_Katsnelson06,Yazyev_Katsnelson08}}. 
\textcolor{black}{Together with the large MAE, it also} gives an exceptionally strong preference for an out-of-plane ferromagnetic alignment (\textcolor{black}{with a large Curie temperature}) of the antisite magnetic moments,  thus creating optimal conditions for also opening a gap in the DSS \cite{Nomura_Nagaosa11, Checkelsky_Tokura12, Tokura_Tsukazaki19}.

To further understand the antisite-induced magnetic state, we analyze its spatial properties in the 6\% Bi$_\mathrm{Se_{1,2}}$ systems. In \Fig{f2}(c) we resolve the spin magnetic moments into layers of thickness $h$ equal to the Bi-Se bond-length projected onto the c-axis, see black box in \Fig{f2}(d). We find that magnetism is only present in the surface layers, with a peak in the layer of the defect. For the in-plane behavior we show in \Fig{f2}(d) the real-space magnetization density of the surface atoms for the Bi$_\mathrm{Se_{1}}$ defect. We find that the magnetization is semi-itinerant, extending with a three-fold spatial pattern from the defect to distances well beyond the primary unit-cell. 
To quantify the itinerancy, we study how the magnetization accumulates with distance away from the antisite defect. For this we plot in \Fig{f2}(e) the net magnetization within a rhombus with the same shape as the unit-cell and with thickness $h$ and centered around the antisite defect with varying side length $d$. For Bi$_\mathrm{Se_{1}}$ the magnetization continuously increases with $d$, demonstrating semi-itinerancy. However, for the Bi$_\mathrm{Se_{2}}$ defect we find that the magnetization is localized since a plateau develops when $d$ equals about a \textcolor{black}{quarter} of the lattice parameter of the unit-cell in the a-b plane. 

\section{Surface energy gap}
\begin{figure}[]
\begin{center}
\includegraphics[width=0.99 \linewidth]{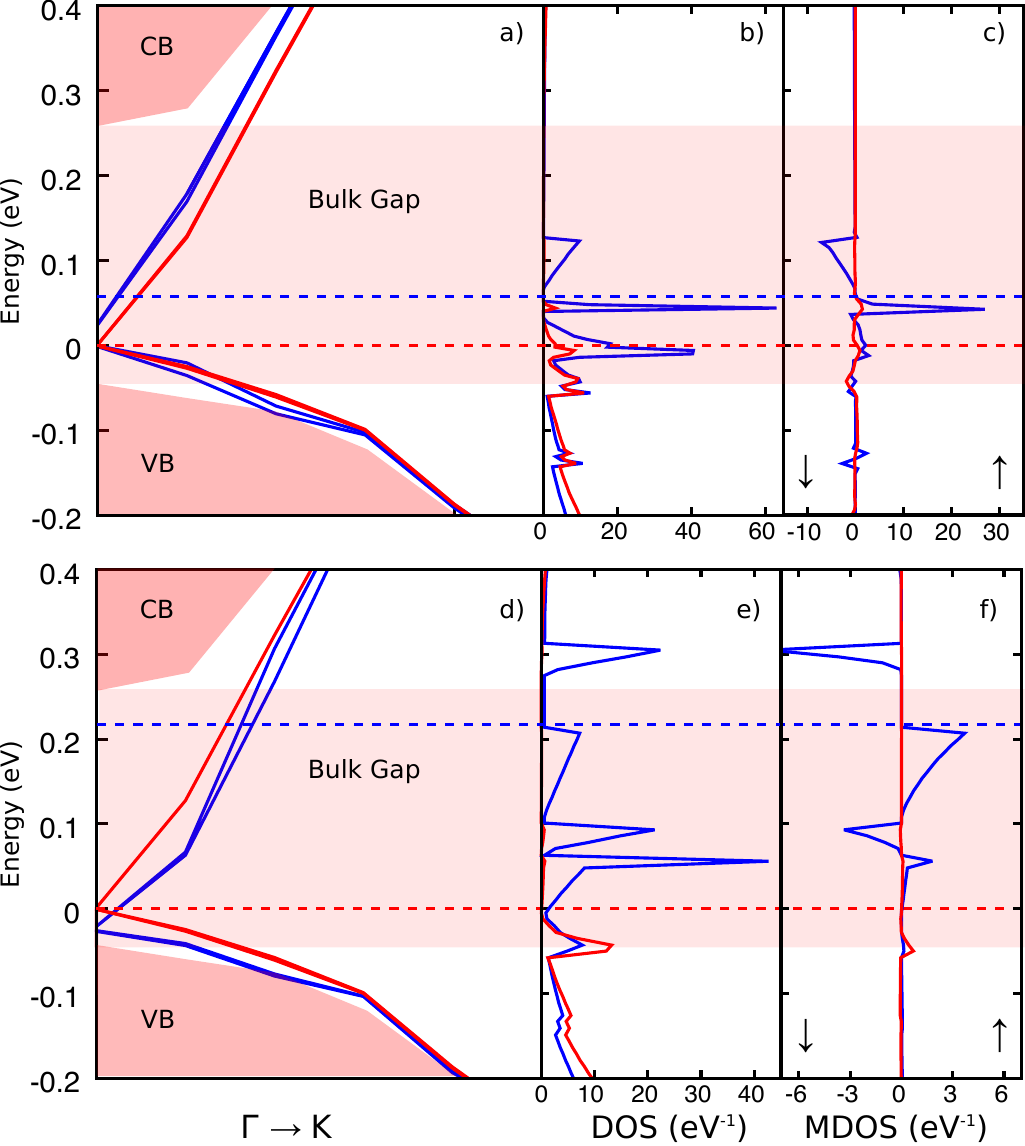}
\end{center}
\caption{(a,d) Band structure along $\Gamma-K$ for pristine (red) and Bi$_\mathrm{Se_1}$ (a) and Bi$_\mathrm{Se_2}$ (d) 6\% defected (blue) TI system with defect states removed. Bulk conduction and valence bands (dark pink) are aligned to create a common bulk band gap (light pink), with the Fermi level (dotted lines) in pristine system set to 0~eV. (b,e) Complete density of states (DOS) and (c,f) magnetization density of states (MDOS) in the bulk (red) and the surface quintuple layer (blue) for the defected systems.}
\label{f3}
\end{figure}

Having established finite magnetism from intrinsically non-magnetic Bi$_\mathrm{Se}$ defects, we turn to investigating the electronic spectrum in detail. Since we find an exceptionally strong MAE, effectively guaranteeing an out-of-plane magnetic moment, we are particularly interested in how the magnetization affects the topologically protected DSSs in Bi$_2$Se$_3$. 
In \Fig{f3}(a) we plot the band dispersion along the $\Gamma$-K direction for the Bi$_\mathrm{Se_1}$ defect system at 6\% concentration (blue) and compare it with the equivalent but defect-free system (red). 
To be able to effectively compare the two systems, we first set the Fermi level of the pristine slab to 0 eV at the Dirac point. We then align the spectrum of the defected slab such that the valence (VB) and conduction (CB) bands perfectly align for the defect and defect-free systems, see SM \cite{SM}. This is possible since both systems reach bulk conditions in the interior of the slabs. We refrain from plotting all bands belonging to the bulk but instead conceptually show their extent in the dark pink regions. We see a clear bulk band gap ranging from -0.05 eV to 0.27 eV (light pink), in agreement with previous predictions\cite{Zhang_Zhang09}. We also identify an intrinsic doping produced by the antisite defect, as the  Fermi level (dotted lines) of the antisite defect system falls at a slightly higher energy (61 meV with 6\% defects). 

We next focus on the in-gap region, where we expect to find the DSS, but also defect states generated by the antisite defects. Initially we are interested only in the intrinsic DSSs, and therefore first exclude all bands belonging to the antisite defects. We can do this easily beyond the $\Gamma$ point, as there the DSS and the antisite defect states have very different orbital and spatial characters: states belonging to the DSS is present throughout the surface, while the antisite defect states are heavily localized at the defect. At the $\Gamma$ point we find finite hybridization between the DSS and some defect states, but, nonetheless, we can still remove the defect states based on their orbital weights and flat energy dispersion (due to their localization), see SM \cite{SM}.
In this way we extract and plot only the DSSs for the Bi$_\mathrm{Se_1}$ defect in ~\Fig{f3}(a). The DSS in the pristine system (red) and antisite system (blue) are very similar at higher energies, both showing a linear Dirac spectrum with the same slope. However, at low energies we find a clear 24~meV energy gap induced in the antisite system. The gap size at the $\Gamma$ point is fully consistent with the slope of the DSS at higher energies. In \Fig{f3}(d) we plot the equivalent bands for the Bi$_\mathrm{Se_2}$ defect, but here the DSS energy gap is negligible, despite the finite magnetization. This is another property, along with the spatial extent of the magnetization, where we observe contrasting behavior for Bi$_\mathrm{Se_1}$ and Bi$_\mathrm{Se_2}$ defects.

\section{Density of states}
To gain further insight into the magnetization and DSS energy gap we investigate the density of states (DOS). 
In ~\Fig{f3}(b,e) we compare the DOS in the bulk (red) with that of the surface quintuple layer of the Bi$_\mathrm{Se_1}$ and Bi$_\mathrm{Se_2}$ antisite systems (blue), respectively. By comparing these two DOS, we find that the DOS predominantly belonging to the Bi$_\mathrm{Se_{1(2)}}$ antisite defect occupy an energy window ranging from around -20 (0) meV  to 120 (220) meV, with respect to the pristine Dirac point, thus filling a large part of the bulk energy gap for both types of defects. However, we also observe that the defect states co-exist (in energy) with the induced energy gap for the Bi$_\mathrm{Se_1}$ defect, while for the Bi$_\mathrm{Se_2}$ defect, the defect states are mainly around 60 meV above the Dirac point of that defected system. By additionally studying the orbital character of all low-energy bands near the $\Gamma$-point, we find that the non-dispersive Bi$_\mathrm{Se_1}$ states overlapping in energy with the Dirac point strongly hybridizes with the DSS, see SM \cite{SM}. This hybridization explains why the Bi$_\mathrm{Se_1}$ defect both generates a semi-itinerant magnetization and opens an energy gap in the DSS by effectively acting as a TRS breaking perturbation on the DSS. The strong hybridization between the magnetic Bi$_\mathrm{Se_1}$ defect states and the DSS also provides the necessary pathway for a strong exchange coupling to align the antisite magnetic moments \cite{Mahani_Canali14}. Here, with Bi$_\mathrm{Se}$ being an inherent defect, it naturally has the same ($s$,$p$) orbital character as the DSS. {\color{black} This provides an additional clear advantage, beyond the energy overlap, in generating a strong exchange coupling over transition metal atoms with their $d$ orbital character  \cite{Islam_Sessi18}.}

The semi-itinerant magnetism and finite energy gap for the Bi$_\mathrm{Se_1}$ system should be contrasted with the behavior of the Bi$_\mathrm{Se_2}$ system. While the Bi$_\mathrm{Se_2}$ defect states have a finite magnetization, they have a negligible overlap in energy with the DSS around its Dirac point. As a consequence, they do not effectively couple to the DSS and thus the magnetization stay localized and the energy gap in the DSS remains vanishingly small. Thus we conclude that a mere presence of an out-of-plane magnetic defect moments does not guarantee the opening up of an energy gap in TIs, but that an effective coupling between the magnetic defect states and the DSS needs to be present as well. 

The creation of in-gap defect-induced resonance states for strong potential defects has previously been established for generic 2D Dirac materials \cite{Wehling_Balatsky14}, including the DSS in TIs \cite{Black-Schaffer_Balatsky12a} and also in the presence of finite magnetic moments \cite{Black-Schaffer_Balatsky15}. Our ab-initio results establish that naturally occurring surface antisite Bi$_\mathrm{Se}$ defects act as such strong potential scatterers, inducing in-gap resonance states. This then also implies a so-called two-fluid behavior \cite{Sessi_Balatsky16}, with both the dispersive DSS and the localized impurity resonance states filling the TI bulk energy gap, as is clearly visible in \Fig{f3}.

Finally, we compare the magnetization density of states (MDOS) between the bulk and surface in \Fig{f3}(c,f) for the Bi$_\mathrm{Se_{1,2}}$ defects. We find that the magnetization in the system is almost exclusively associated with the in-gap defect states. This also finally offers an explanation as to why the antisite defect states spontaneously become magnetized in the first place: the defect resonance states generate a large DOS at the Fermi level $\rho(E_F)$, which necessarily becomes prone to spontaneous magnetization. In its simplest incarnation the instability towards magnetism is given by a Stoner-like criterion $\rho(E_F) U>1$, where $U$ is the electron-electron interaction strength \cite{Black-Schaffer_Yudin14}. Our ab-initio results on antisite Bi$_\mathrm{Se}$ defects on the surface of the TI Bi$_2$Se$_3$ show that antisite defects are indeed strong enough potential scatterers to generate these low-energy defect-induced resonances, which then thanks to finite exchange interactions in the TI also become spontaneously magnetized, see SM for further information \cite{SM}. {\color{black} Moreover, naturally occurring Se vacancy states will not be detrimental as they appear at a larger energy \cite{Black-Schaffer12b,Mann_Shih13}}

\section{Conclusions}
Our fully relativistic ab-initio calculations show that intrinsic antisite Bi$_\mathrm{Se}$ defects in the surface layer of the TI Bi$_2$Se$_3$ generates a finite energy gap in the topologically protected DSSs. 
The antisite defect produces low-energy resonance states, which spontaneously become magnetic with an exceptionally large MAE guaranteeing an out-of-plane magnetic moment. With the defect states also overlapping in energy with the Dirac point, they hybridize with the DSS and thus the surface antisite defect acts as an effective magnetic field opening an energy gap in the DSS. For antisite defects buried in the first subsurface layer we also find a finite magnetization, but the overlap with the DSS is negligible and thus no measurable energy gap is present in the DSS. These results illustrates both that naturally occurring defects can produce a magnetic TI and that magnetic defect moments require effective coupling to the DSS to open an energy gap. Moreover, the results provide an important observation on the site dependence of defects in exhibiting essential physics of TIs.

We thank A.~Bouhon, X.~Chen, R.~Esteban-Puyuelo, B.~Ghosh, M.~Mashkoori, F.~Parhizgar,  and D.~Wang for fruitful discussions. This work was supported by the Swedish Research Council (Vetenskapsr\aa det, Grant Nos.~2014-3721 and 2018-03488), the Swedish Foundation for Strategic Research (SSF), and the Wallenberg Academy Fellows program through the Knut and Alice Wallenberg Foundation. The simulations were performed on resources provided by the Swedish National Infrastructure for Computing (SNIC)
at HPC2N, NSC, PDC, and UPPMAX.

\bibliography{newref}
\end{document}


\title{Supplementary material for ``Spontaneous ferromagnetism and finite surface energy gap in the topological insulator Bi$_2$Se$_3$ from surface Bi$_\mathrm{Se}$ antisite defects''}

\author{Suhas Nahas}
\author{Biplab Sanyal}
\author{Annica M. Black-Schaffer}
\affiliation{Department of Physics and Astronomy, Uppsala University, Box 516, SE-751 20 Uppsala, Sweden}
\maketitle

In this Supplementary material we provide additional information supporting the conclusions drawn in the main text. We start by showing the structural distortions around the  Bi$_\mathrm{Se_2}$ anti-site defects in the surface and the bulk. \textcolor{black}{Then, we provide supporting information on the interactions between two anti-site defects.} This is followed by showing how we align the defected and pristine band structures. Then, we show how we can extract the Dirac surface states (DSSs) for the Bi$_\mathrm{Se_{1,2}}$ antisite defect slabs based on the orbital character of the antisite defect states. \textcolor{black}{Finally, we show how the magnetization density of states varies for different defect concentration and how this is consistent with a Stoner mechanism for magnetization.}

\begin{figure}[H]
\begin{center}
\includegraphics[width=0.5 \linewidth]{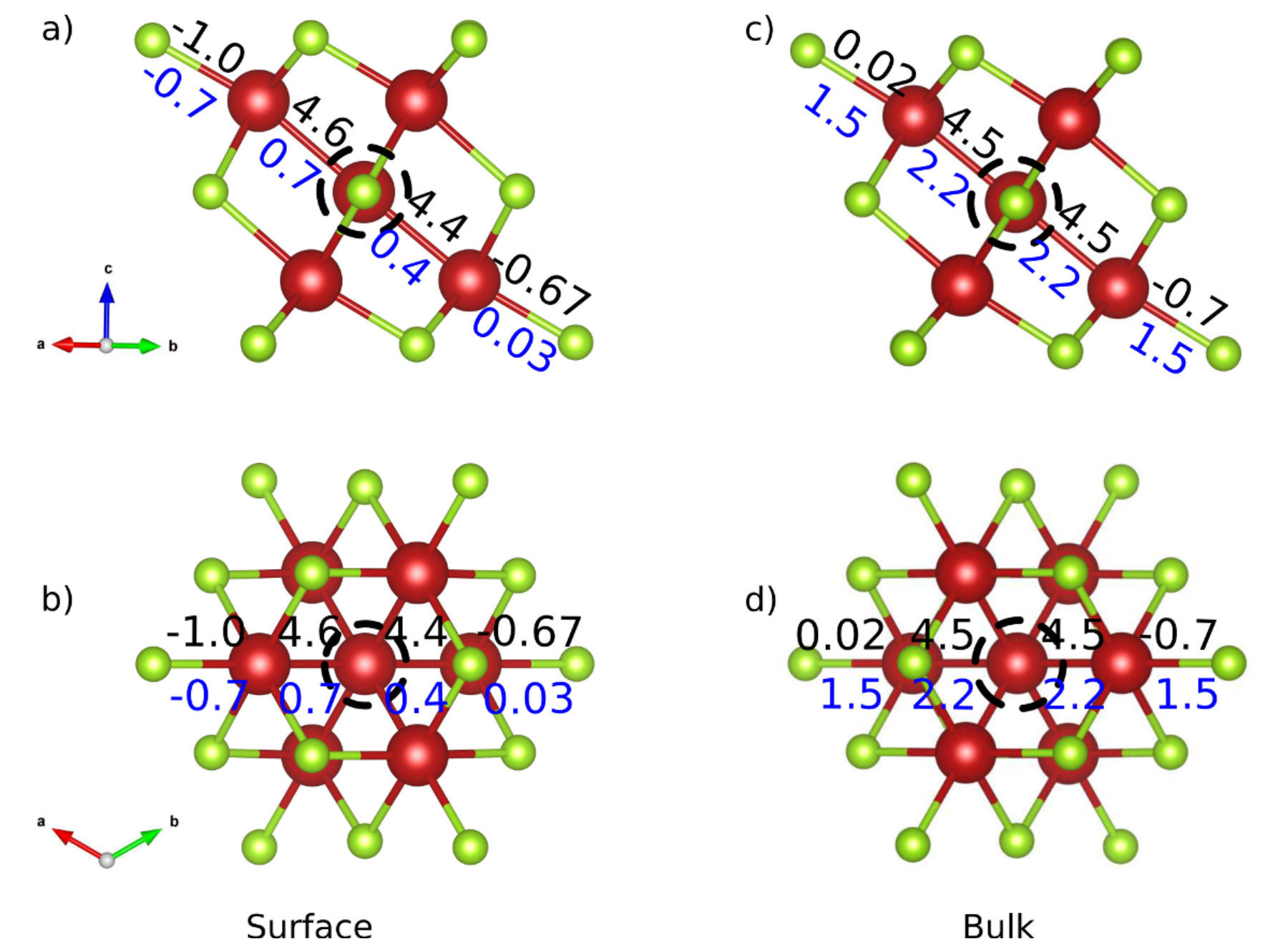}
\caption{Side a (c) and top b (d) views of the neighborhood of the Bi$_\mathrm{Se_2}$ antisite defect in the surface (bulk) of Bi$_2$Se${_3}$. The total relative (in \%) bond length change in the presence (absence) of spin-orbit coupling in shown in black (blue).}
\label{f1}
\end{center}
\end{figure}

\begin{center}
\section{Structural distortions for Bi$_\mathrm{Se_2}$ in the surface and bulk}  
\end{center}
In the main text, we show that spin-orbit coupling greatly influences the local atomic structure of the Bi$_\mathrm{Se_1}$ defect by displaying the relative change in the bond lengths close to the antisite defect, both in the presence (black) and absence (blue) of spin-orbit coupling. Here in \fig{f1} we show that the same behavior for the Bi$_\mathrm{Se_2}$ defect, both when it is in the surface as well as in the bulk. As seen, there are similar structural distortions locally around antisite defects for all layers. Notably, including spin-orbit coupling is very important in determining the correct atomic structure.

{\color{black}
\begin{center}
\section{Interaction between anti-site defects} 
\end{center}

In the main text we discussed in detail the spontaneous (ferro)magnetization and out-of-plane anisotropy for the  case of 6\% Bi$_\mathrm{Se_1}$. We did this by using a supercell approach with only one Bi$_\mathrm{Se}$ antisite defect per supercell. However, this does not fully capture the pair-wise interaction between two Bi$_\mathrm{Se_1}$, as all the Bi$_\mathrm{Se_1}$ in such a system are constrained by the periodic boundary conditions (PBC) to have the same magnetization.

Here, we offer complementary data on two antisite defects in a single supercell, keeping the concentration at $\sim$ 6\%, to confirm that the results reported in the main text regarding the spontaneous magnetization and the MAE holds true even when pairwise interactions between antisite defects are considered. In this setup the defects are at a distance of $14.5 {\AA}$, as shown in \fig{f2}, and the antisite defects are now allowed to have magnetic moments independent of one another.

	 We find that a net out-of-plane magnetization also exist in this case, with each Bi$_\mathrm{Se_1}$ having a moment of $\sim 1\mu_B$. \Fig{f2} shows the spatial magnetization density and it is clearly seen to be consistent with the semi-itinerant magnetism reported in Fig.~2(d) in the main text. We also study whether the MAE is affected due to the inclusion of pair-wise interactions between Bi$_\mathrm{Se_1}$. By using the same method as in the main text, we calculate the MAE per antisite defect and find it to be $\sim$ 15 meV. This again is in close agreement with the values reported in the main text. Further, we use this setup to calculate the exchange coupling as the energy difference between the parallel (FM) and anti-parallel (AFM) configurations of the magnetic moments and find it to be 3.6 meV, as reported in the main text.

\begin{figure}[H]
\begin{center}
\includegraphics[width=0.9 \linewidth]{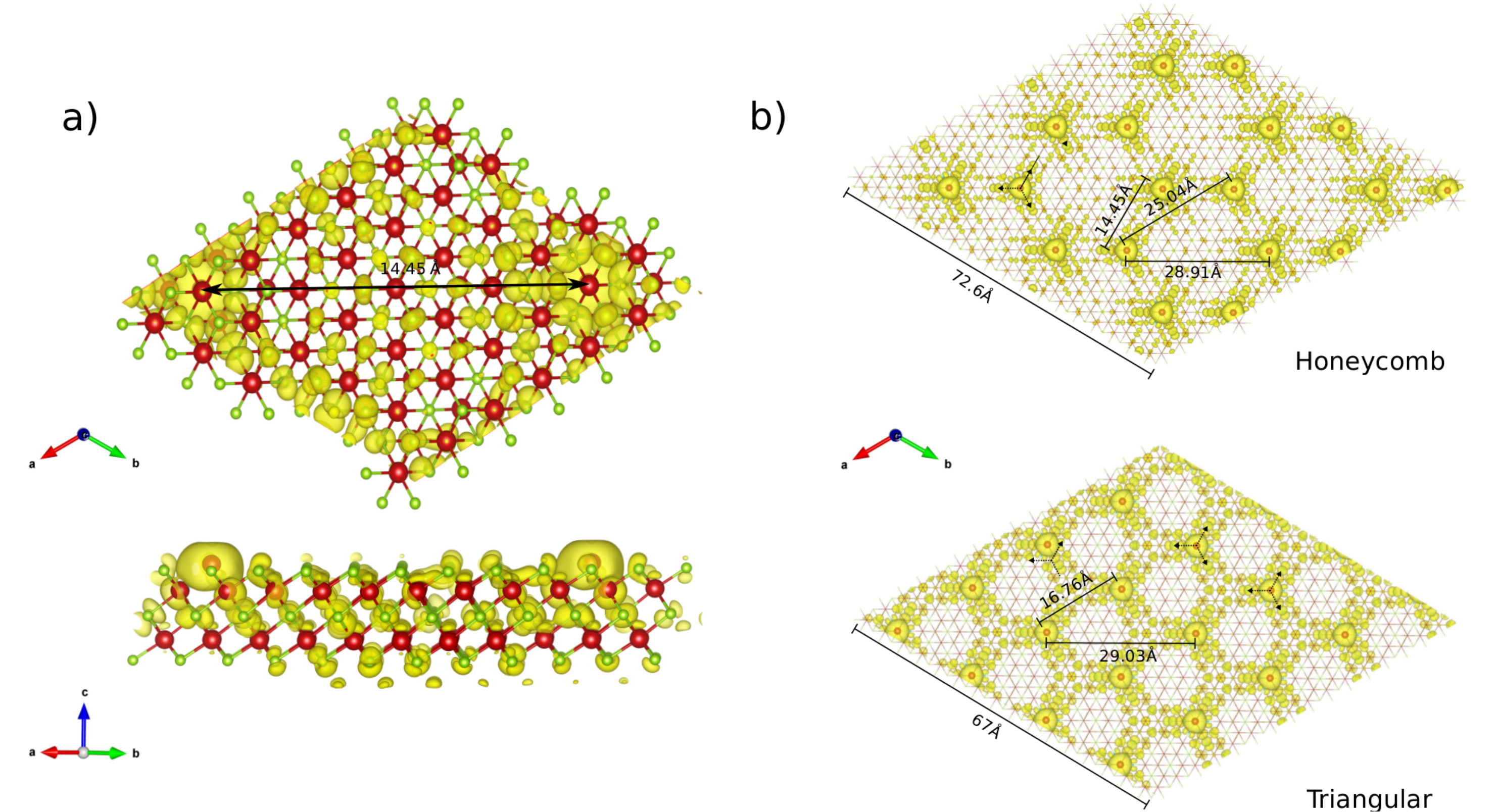}
\caption{(a) Real space magnetization density (iso-value 0.1 times the maximum) in a supercell with two Bi$_{\mathrm{Se_1}}$ defects at 6\% concentration, with top (top) and side (bottom) views. (b) Triangular (top) and honeycomb (bottom) arrangements of Bi$_{\mathrm{Se}_1}$ antisite defects and their corresponding real-space magnetization densities. The three-fold (clover-leaf) magnetization for each Bi$_{\mathrm{Se}_1}$, is schematically represented by three outward arrows.}
\label{f2}
\end{center}
\end{figure}

	We also comment on the atomic arrangement of the antisite defects. For the case of two Bi$_\mathrm{Se_1}$ in the supercell, discussed here in the Supplementary material, the antisite defects are arranged in a honeycomb lattice, as shown in \fig{f2}(b) (top). In contrast, for the case discussed in the main text, i.e.~one Bi$_\mathrm{Se_1}$ per supercell, the antisite defects are arranged in a triangular lattice as shown in \fig{f2}(b) (bottom). Despite the differences, we find that the spontaneous magnetization and the MAE are similar, irrespective of the arrangements of the Bi$_\mathrm{Se_1}$ defects.  To illustrate the spatial properties more clearly, we represent the spread of the three-fold (clover-leaf) magnetization with three out-ward arrows in \fig{f2}(b). We find that in the honeycomb arrangement the spread of the magnetization is such that the arrows belonging to one Bi$_\mathrm{Se_1}$ points in the direction of the other Bi$_\mathrm{Se_1}$. While in the triangular case, the magnetization of each Bi$_\mathrm{Se_1}$  extends in such a way that the  arrows point towards the center of the triangle formed by the antisite lattice. Based on these results, we conclude that the magnetization survives irrespective of its relative (between two Bi$_\mathrm{Se_1}$ ) spread and hence should also not depend on the detailed arrangement on the defects, as long as the concentration of Bi$_\mathrm{Se_1}$ remains the same. This notably should make our results and conclusions insensitive to disorder effects.
}

\begin{center}
\section{Aligning the bands of defected and pristine Bi$_2$Se${_3}$} 
\end{center}

\begin{figure}[H]
\begin{center}
\includegraphics[width=0.35 \linewidth]{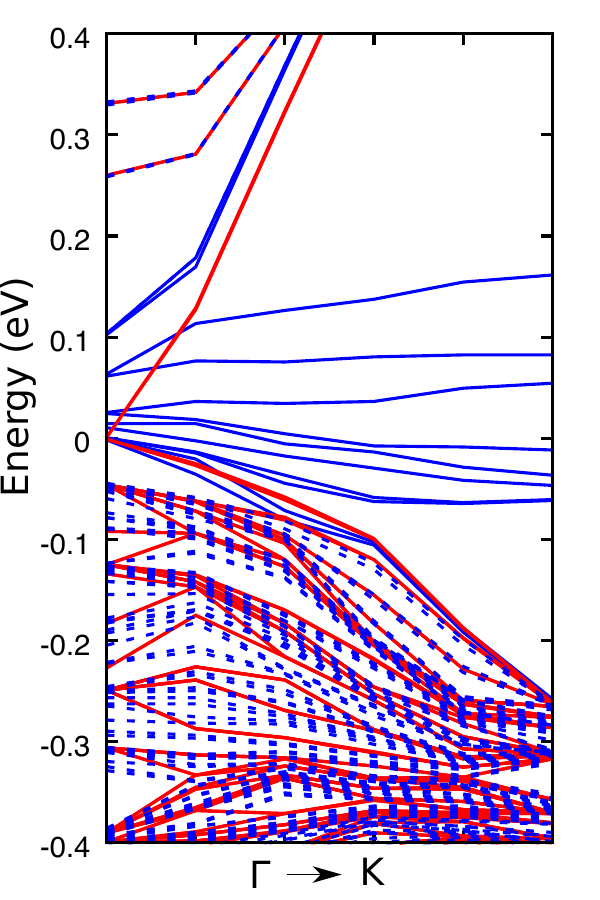}
\caption{Full band structure of the pristine (red) and defected (blue) slab with Bi$_\mathrm{Se_1}$ antisite defects. For the defected system, bulk bands are displayed with dashed lines.
We align the pristine and defected bands such that the valence and conduction bands lie on top of each other. 
}
\label{f4}
\end{center}
\end{figure}

 In order to draw conclusions about the band structure of the antisite Bi$_2$Se${_3}$ we need to properly align it with that of the pristine Bi$_2$Se${_3}$. In \fig{f4} we plot the full band structure  of both the pristine (red) and antisite (blue) slab following the alignment procedure described in the main text. For the defected system we plot the bands belonging to the bulk valence and conduction bands with dotted lines, while the states (defect and DSS) within the gap are shown with solid lines. We can easily make this distinction by noting their spatial extent within the slab.
 Clearly, we see that both the conduction and valence bands of both systems overlap essentially perfectly. This helps us to identify both the DSS and the defect states in the gap region.  

\begin{center}
\section{Extracting the DSS for Bi$_{\mathrm Se_1}$ defects} 
\end{center}

\begin{figure}[H]
\begin{center}
\includegraphics[width=0.55 \linewidth]{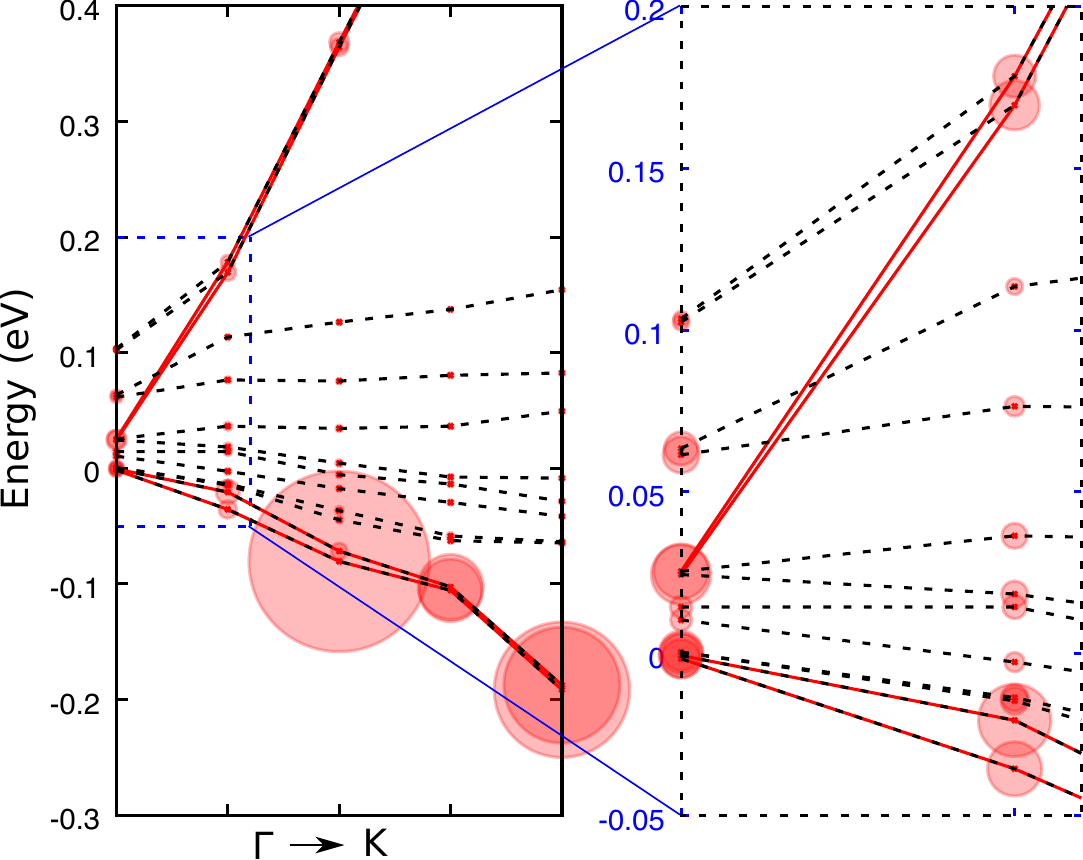}
\caption{Low-energy band structure of the Bi$_\mathrm{Se_1}$ antisite slab. The bands obtained by connecting the energy eigenvalues in the a simple ascending order are shown in black-dotted lines. The diameter of the circles at each eigenvalue is set by the inverse of the orbital weight around the Bi$_{Se_1}$ defect (see text). Larger circles thus correspond to less defect weight. The DSS bands (red) are obtained by connecting the largest circles at each $k$-point. To the right is an enlarged view of the region around the $\Gamma$-point, clearly displaying the energy gap and the finite hybridization between DSS and defect states.}
\label{f5}
\end{center}
\end{figure}

Note that away from the $\Gamma$-point, we find that the DSS and the defect states are both well separated in energy and display large differences in circle diameters, demonstrating a very clear separation between the DSS and localized defect states. Closer the $\Gamma$-point we observe a notable hybridization between the DSS and the defect states. Here, circle diameters for the DSS states are only around three times larger than those of the defect states. Still, this difference makes it possible for us to unequivocally identify a finite energy gap, also consistent with the slope of the DSS. The hybridization in turn explains how the magnetization of the defect states can open up the energy gap as that provides the necessary coupling between the defect states and the DSS.
In order to extract the DSS in the Bi$_{Se_1}$ slab, we study the orbital weights for each low-energy eigenstate. For the purpose of displaying the data we sum the orbital weights for all atoms up to the next-neighbors (within a distance of  5~\AA) of the Bi$_{Se_1}$ defect. Then we use a circle at each energy eigenvalue, whose diameter is set by the inverse of this summed orbital weight, to display how much that particular eigenstate is delocalized over the whole surface, i.e., the larger the circle, the less localized defect character in that eigenstate. Doing this for all in-gap states, we can then connect those eigenvalues that have the largest circles to give the DSS as that is the state that is fully delocalized over the surface. The remaining bands are then identified as localized defect states. In \fig{f5} we show the result, including a zoom-in around the $\Gamma$-point. We clearly observe the lower and upper branch of the DSS as they have significantly larger circles than the remaining states. They also show the characteristic linear dispersion as expected for a DSS.

\begin{center}
\section{Extracting the DSS for Bi$_{\mathrm Se_2}$ defects} 
\end{center}

\begin{figure}[H]
\begin{center}
\includegraphics[width=0.3 \linewidth]{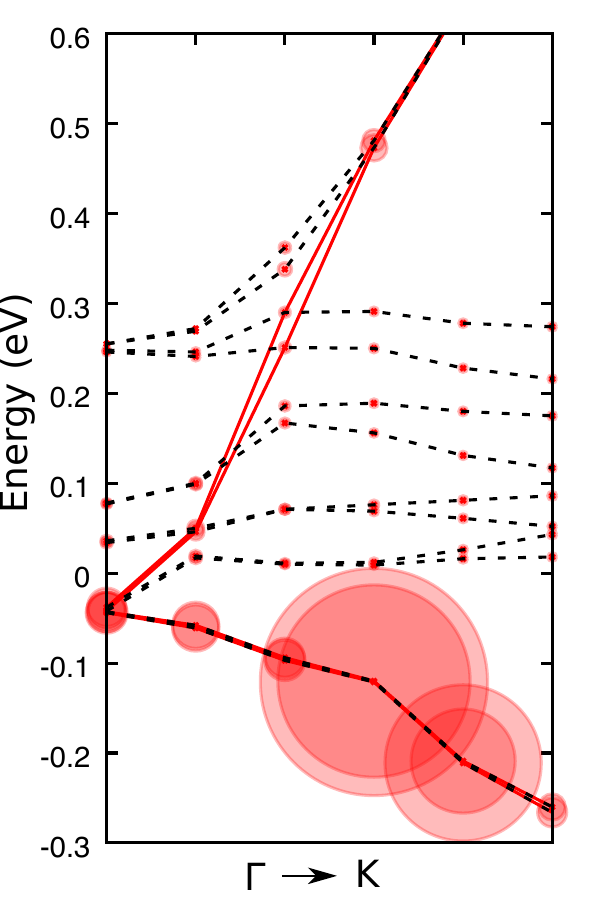}
\caption{Low-energy band structure of the Bi$_\mathrm{Se_2}$ antisite slab. The bands obtained by connecting the energy eigenvalues in the a simple ascending order are shown in black-dotted lines. The diameter of the circles at each eigenvalue is set by the inverse of the orbital weight around the Bi$_{Se_2}$ defect (see text). Larger circles thus correspond to less defect weight. The DSS bands (red) are obtained by connecting the largest circles at each $k$-point.}
\label{f6}
\end{center}
\end{figure}

We follow the same procedure as in the previous section also for the Bi$_{\mathrm Se_2}$ antisite defect slab and show the results in \fig{f6}. While there are large overall similarities with the Bi$_{\mathrm Se_1}$ system, we here find that the defect states are spread to higher energies. In particular, near the $\Gamma$-point, we find that the DSS do not hybridize with the defect states, simply because there are no defect states there. As a consequence, we also find no energy gap in the DSS. Far away from the $\Gamma$-point we however find some minor hybridization between the DSS and defect states, although that does not influence the properties around the Dirac point.

{\color{black}
\begin{center}
\section{Comparing the (M)DOS of 6\% and 25\% concentrations and the Stoner mechanism} 
\end{center}

	In the main text we attribute the spontaneous out-of-plane magnetization to the Stoner instability of the resonance peak near the Fermi level associated with Bi$_\mathrm{Se}$. At the same time, in \fig{f2}(a,b) in the main text, we find that the magnetization and MAE show strong dependences on the concentration. Here, we clarify these differences in behavior by comparing the resonance peaks for the antisite concentrations 6\% and 25\%. For this, we focus on how the peak width and the Stoner exchange splitting change with concentration. 
	
	In \fig{f3} a (b) we show the DOS (MDOS) for both 6\% and 25\% concentrations of Bi$_\mathrm{Se_1}$ antisite defects. In order to look at how the peak width vary with concentration, we calculate the full width at half maxima (FWHM) of the resonance peaks find it to be 4.6 meV and 9.8 meV for 6\% and 25\% concentrations, respectively. Thus, we find that even for a significant increase in the concentration (from 6\% to 25\% ), the peak-width only increases by a factor of 2. It is also interesting to note that the peak width is significantly smaller compared to other systems with spontaneous magnetization.\cite{Tang_Zhang19,Yazyev_Helm07} This small peak width facilitates stability of the magnetic moments at high temperatures.\cite{Edwards_Katsnelson06}
However, we find that the Stoner exchange splitting is much more affected, see \fig{f3}(b) with an increase of the concentration from 6\% to 25\%. For the 6\% case, we find an exchange splitting of 80 meV (strong ferromagnet), while for 25\% we see that the splitting is just 6 meV (weak ferromagnet). Thus, we conclude that the exchange splitting depends strongly on the concentration, while the peak width is less affected. This clearly illustrates that, while both systems spontaneously magnetize, they show rather different magnetization properties. 

Using the exchange splitting ($E_\mathrm{ex}$) and the net magnetization ($m$) we can also roughly estimate\cite{Mag} the Coulumb interaction ($U$) by using the expression

\begin{gather}
U=\frac{E_\mathrm{ex}}{m}
\end{gather}

Thus, for 6\% Bi$_\mathrm{Se}$ (which is the main focus in main text), we get $U = 93$~meV using this expression.


\begin{figure}[H]
\begin{center}
\includegraphics[width=0.5 \linewidth]{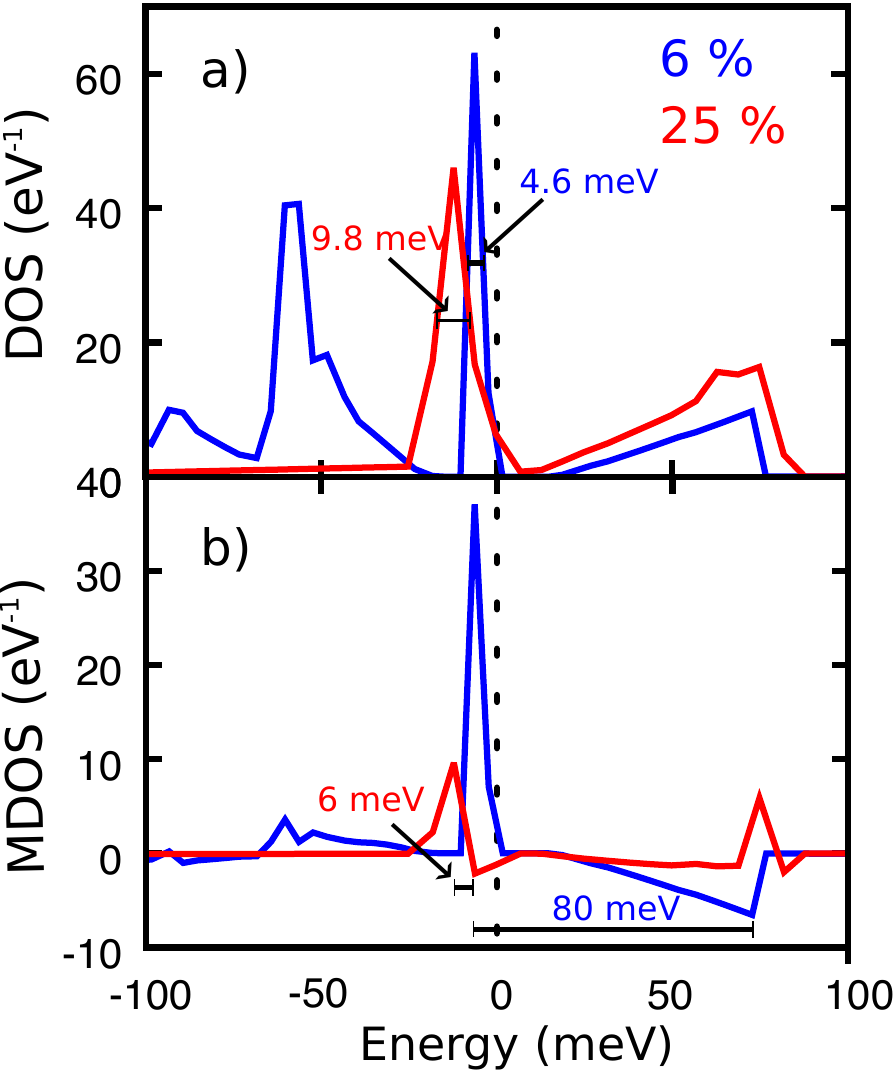}
\caption{a) DOS near the Fermi level for 6\% (blue) and 25\% (red) concentration of antisite defects Bi$_\mathrm{Se_1}$, with full width at half maxima (FWWHM) indicated. b) MDOS near the Fermi level for 6\% (blue) and 25\% (red) concentration of antisite defects Bi$_\mathrm{Se_1}$, with the Stoner exchange splitting energy indicated.}
\label{f3}
\end{center}
\end{figure}

}

\bibliography{newrefS}